\documentclass[fleqn,twoside]{article}
\usepackage{espcrc2}
\usepackage{axodraw}
\usepackage{color}

\usepackage{graphicx}
\usepackage[figuresright]{rotating}

\definecolor{Yellow}{cmyk}{0,0,1,0}   %
\definecolor{Red}{cmyk}{0,1,1,0}   %
\definecolor{Magenta}{cmyk}{0,1,0,0}   %
\definecolor{Blue}{cmyk}{1,1,0,0}   %
\definecolor{Cyan}{cmyk}{1,0,0,0}   %
\definecolor{Green}{cmyk}{1,0,1,0}   %
\definecolor{Brown}{cmyk}{0,0.81,1,0.20}   %
\newcommand{\met}{\rlap{\,/}E_T}

\newcommand{\AmS}{{\protect\the\textfont2
  A\kern-.1667em\lower.5ex\hbox{M}\kern-.125emS}}

\title{Particle Dark Matter from Physics Beyond the Standard Model}

\author{Konstantin Matchev\address[]{Physics Department,
        University of Florida, Gainesville FL 32611, USA}%
        \thanks{Present address: 
                Institute for High Energy Phenomenology,
                Newman Laboratory for Elementary Particle Physics,
                Cornell University, Ithaca, NY 14853, USA.
		Work supported in part by the US Department of Energy 
                under grant DE-FG02-97ER41029.}}

\begin{document}

\begin{abstract}
In this talk I contrast three different particle dark matter candidates,
all motivated by new physics beyond the Standard Model:
supersymmetric dark matter, Kaluza-Klein dark matter, and scalar dark matter.
I then discuss the prospects for their discovery and identification in both
direct detection as well as collider experiments.
\vspace{1pc}
\end{abstract}

\maketitle

\section{DARK MATTER AND PHYSICS BEYOND THE STANDARD MODEL}

So far the existence of the dark matter (DM) is our best 
{\em experimental} evidence for new physics 
beyond the Standard Model. The most recent WMAP data
\cite{Bennett:2003bz}
confirm the standard cosmological model 
and precisely pin down the amount of cold dark matter
as $0.094<\Omega_{CDM}h^2<0.129$, in accord with 
earlier indications.
Nevertheless, the exact nature of the dark matter
still remains a mystery, as all
known particles are excluded as DM candidates.
This makes the dark matter problem the most pressing
phenomenological motivation for particles and interactions
beyond the Standard Model (BSM).

The weakly interacting massive particles (WIMPs) 
represent a whole class of particle dark matter candidates
which are well motivated by both particle physics and astrophysics.
On the particle physics side, many BSM theories predict stable WIMPs.
On the astrophysics side, the result from the calculation of the WIMP
thermal relic density falls in the right ballpark.
WIMPs also offer excellent opportunities for detection:
since they must have been able to annihilate sufficiently fast
in the Early Universe, they should also produce observable signals
in direct and indirect dark matter detection experiments.

Generally speaking, it is rather easy to cook up BSM dark matter.
The main steps in model-building are the following:
\begin{enumerate}
\item enlarging the particle content of the SM;
\item introducing a symmetry which guarantees 
that one of the newly introduced particles is stable;
\item fudging the model parameters until
the lightest new stable particle is neutral
and has the proper thermal relic density.
\end{enumerate}
In the next three sections I will illustrate this 
model-building recipe with three generic examples
of BSM theories with good DM candidates.
Section~\ref{sec:susy} is devoted to supersymmetry~\cite{Chung:2003fi}
in its most popular version (minimal supergravity),
where the DM candidate is the lightest superpartner.
In Section~\ref{sec:ued} I consider the model of 
universal extra dimensions~\cite{Appelquist:2000nn} 
where the lightest Kaluza-Klein mode is the DM particle.
Then in Section~\ref{sec:LH} I discuss DM in
Little Higgs theories~\cite{Arkani-Hamed:2001nc}. 
In each case, after a 
brief introduction to the theory I will 
discuss why the DM particle is stable,
what is its preferred mass range
and what are the discovery prospects for its direct
detection. The potential discovery and
discrimination of these alternatives at 
high energy colliders
is the subject of Section~\ref{sec:colliders}
while Section~\ref{sec:conclusions} contains
a comparative summary and conclusions.
The subject of distinguishing these scenarios in
astroparticle physics experiments in space is 
covered in J. Feng's contribution to these proceedings.

\section{SUSY DARK MATTER}
\label{sec:susy}

Supersymmetry (SUSY) is a theory with extra dimensions described by
new {\em anticommuting} coordinates $\theta_\alpha$. The theory is
defined in terms of superfields living in {\em superspace} $\{x^\mu,\theta\}$:
\begin{equation}
\Phi(x^\mu,\theta) = \phi(x^\mu) + \psi^\alpha(x^\mu)\theta_\alpha
+ F(x^\mu) \theta^\alpha\theta_\alpha\ ,
\end{equation}
where the pair $\{\phi, \psi\}$ is a SM particle and its superpartner.
Supersymmetry thus predicts a host of new particles, offering a
potential solution to the DM problem.
A robust prediction of SUSY is that 
each superpartner has couplings identical to
and spin differing by $1/2$ from the corresponding 
SM particle. Unfortunately, the superpartner 
mass spectrum is very model dependent as it is related 
to the details of SUSY breaking.

All superpartners are charged under a discrete symmetry
called $R$-parity, which is a remnant of translational 
invariance along the $\theta$ coordinates.
The $R$-parity assignments are
$+1$ for SM particles and $-1$ for their
superpartners. In many models one requires
that $R$-parity is conserved, which provides 
a simultaneous solution to the proton decay problem.
As a bonus, $R$-parity conservation guarantees that
virtual supersymmetric contributions
only arise at the loop level, thus avoiding constraints 
from precision electroweak data.
As a result of $R$-parity conservation, 
the lightest superpartner (LSP), being
odd under $R$-parity, is absolutely stable and 
becomes a DM candidate. In the remainder of this Section,
we shall restrict ourselves to the case of minimal
supergravity (mSUGRA)
and describe the cosmologically preferred
parameter space with a WIMP LSP.

\subsection{The LSP as a SUSY WIMP}

\begin{figure}[tb]
\includegraphics[width=17pc]{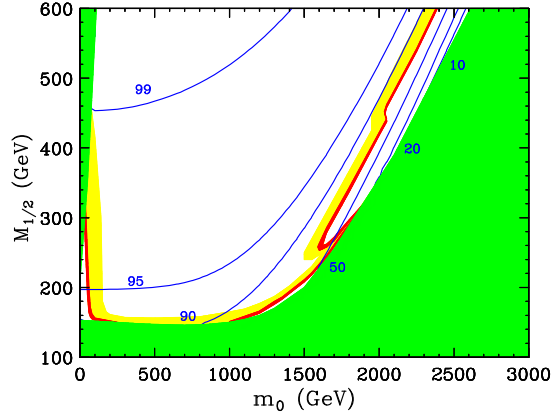}
%\framebox[55mm]{\rule[-21mm]{0mm}{43mm}}
\caption{Contours of constant gaugino fraction $R_\chi$ in percent,
for $\tan\beta=10$, $A_0=0$, $\mu>0$ and $m_t=174$ GeV. The (green)
shaded regions are excluded from the requirement that the LSP be neutral (left)
and the chargino mass bound from LEP (bottom and right).
The light (yellow) shaded region in between corresponds to the
pre-WMAP preferred range of $0.1\le\Omega h^2\le0.3$. The red (dark) 
shaded region is the remaining parameter space after WMAP.
From Ref.~\protect\cite{Feng:2000zu}.}
\label{fig:gfraction}
\end{figure}

The desired WIMP in supersymmetry is the
lightest neutralino $\tilde\chi^0_1$, which is a mixture of
the superpartners of the hypercharge gauge boson ($\tilde b^0$), 
the neutral $SU(2)_W$ gauge boson ($\tilde w^0$), 
and the two neutral Higgs bosons ($\tilde h_u^0$, $\tilde h_d^0$):
\begin{equation}
\tilde\chi^0_1 =
  a_1\, \tilde{b}^0 
+ a_2\, \tilde{w}^0 
+ a_3\, \tilde{h}^0_u
+ a_4\, \tilde{h}^0_d\ .
\end{equation}
In large regions of the mSUGRA parameter space, $\tilde\chi^0_1$
indeed turns out to be the LSP. Its gaugino fraction $R_{\chi}$
\begin{equation}
R_{\chi} \equiv |a_1|^2 + |a_2|^2 \approx |a_1|^2
\end{equation}
is shown in Fig.~\ref{fig:gfraction} in the $m_0-M_{1/2}$
slice of the mSUGRA parameter space.
The pre-WMAP (post-WMAP) preferred parameter space is also shown in
yellow (red). We easily see two regions where $\tilde\chi^0_1$
is a good DM candidate. The ``coannihilation'' region at low $m_0$ 
exhibits a gaugino-like LSP, whose relic density is diluted due to
coannihilation processes involving $\tau$ 
sleptons~\cite{Ellis:1998kh}. Alternatively, at large
$m_0$ one finds the so called ``focus point region'' 
\cite{Feng:1999mn}, where the LSP
has a non-negligible higgsino component,
and annihilation is enhanced
due to a complementary set of diagrams 
proportional to gaugino-higgsino mixing~\cite{Feng:2000gh}.
Unlike the coannihilation region, the 
focus point region is characterized by heavy 
scalar superpartners, and as a result 
all virtual supersymmetric effects are suppressed~\cite{Feng:2000bp}.
In the absence of any significant discrepancies
from the SM, the focus point region tends to be
preferred in global fits to the data~\cite{Baer:2003yh}.

\begin{figure}[htb]
\includegraphics[width=17pc]{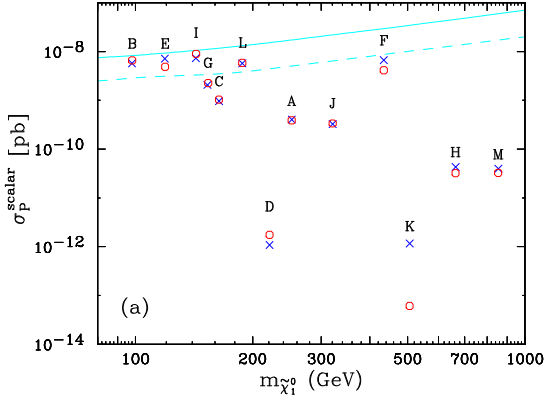}
%\framebox[55mm]{\rule[-21mm]{0mm}{43mm}}
\caption{Elastic cross-sections $\sigma_P^{scalar}$ for spin-independent
$\tilde\chi^0_1$ scattering on protons from two different codes
(circles and crosses) for the 13 benchmark points of
\cite{Battaglia:2001zp}. Projected sensitivities for
CDMS II~\cite{Schnee:gf} %and CRESST~\cite{Bravin:1999fc} 
(solid line) 
and GENIUS~\cite{Klapdor-Kleingrothaus:2000eq} (dashed line) are also
shown. From Ref.~\cite{Ellis:2001hv}.}
\label{fig:scalar}
\end{figure}

\begin{figure}[htb]
\includegraphics[width=17pc]{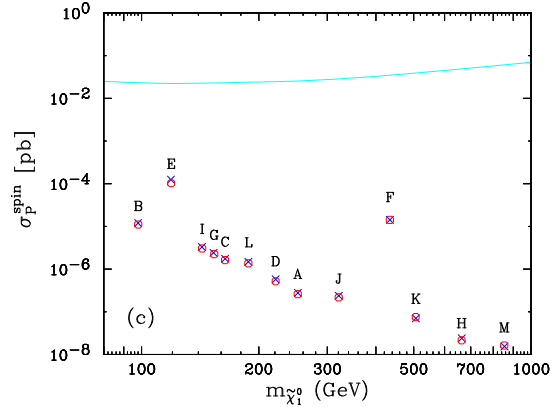}
%\framebox[55mm]{\rule[-21mm]{0mm}{43mm}}
\caption{The same as Fig.~\ref{fig:scalar}, for spin-dependent
$\tilde\chi^0_1$ scattering on protons. The solid line is the
projected sensitivity of a 100 kg NAIAD array~\cite{Spooner:kt}.
From Ref.~\cite{Ellis:2001hv}.}
\label{fig:spin}
\end{figure}

\subsection{Direct detection of SUSY dark matter}

The prospects for direct detection of SUSY WIMPs
in experiments sensitive to spin-independent
(spin-dependent) WIMP-nucleon interactions
are depicted in Fig.~\ref{fig:scalar} (Fig.~\ref{fig:spin})
\cite{Ellis:2001hv},
for the set of 13 SUSY benchmark points from \cite{Battaglia:2001zp}.
We see that spin-independent scattering offers the best possibilities 
for detection -- about half of the benchmark points fall within
the sensitivity range of the upcoming experiments.
Unfortunately, we also see that due to accidental cancellations,
one cannot place a firm lower bound on 
the predicted signal (see, e.g. points D and K).

Regarding spin-dependent scattering, the prospects are less optimistic,
as all 13 benchmark points fall far below the experimental sensitivity.
Therefore, a likely signal in the spin-independent 
channel will still leave open the question of the exact nature of the 
DM particle, and its spin, in particular. In the next two Sections, 
we shall describe alternative DM candidates whose main 
difference from SUSY WIMPs is their spin.

\section{UNIVERSAL EXTRA DIMENSIONS}
\label{sec:ued}

Universal Extra Dimensions~\cite{Appelquist:2000nn}
is an extra dimension theory with new {\em bosonic} coordinates $y$
(in the simplest case of only one extra dimension, $y$ spans a 
circle of radius $R$):
\begin{eqnarray}
\Phi(x^\mu,y) &=& \phi(x^\mu) 
+ \sum^\infty_{i=1} \phi^n(x^\mu)\cos(ny/R) \nonumber \\
&&+ \sum^\infty_{i=1} \chi^n(x^\mu)\sin(ny/R)
\end{eqnarray}
Each SM field $\phi$ ($n=0$) has an infinite tower of Kaluza-Klein (KK)
partners $\phi^n$ and $\chi^n$ with identical spins and couplings
and masses of order $n/R$. In order to obtain chiral fermions,  
the extra dimension is compactified on an $S_1/Z_2$ orbifold.
Fields which are odd under the $Z_2$ orbifold symmetry
do not have zero modes ($\phi=\phi^n=0$), which allows us to 
project out the zero modes of the fermions with
the wrong chiralities and the 5th component of the gauge fields.
The remaining zero modes are just the Standard Model
particles in 3+1 dimensions.

A peculiar feature of UED is the conservation of
KK number at tree level, a simple consequence
of momentum conservation along the extra dimension.
However, bulk and brane radiative effects 
\cite{Georgi:2001ks,Cheng:2002iz}
break KK number down to a discrete conserved quantity,
the so called KK parity: $(-1)^n$, where $n$ is the KK level.
KK parity ensures that the lightest KK partner (LKP) at level one,
being odd under KK parity, is stable and can be a DM candidate.
Similar to the SUSY case, new physics contributions
to various precisely measured low-energy observables 
only arise at the loop level and are small.

\subsection{The LKP as a UED WIMP}

\begin{figure}[tb]
\includegraphics[width=17pc]{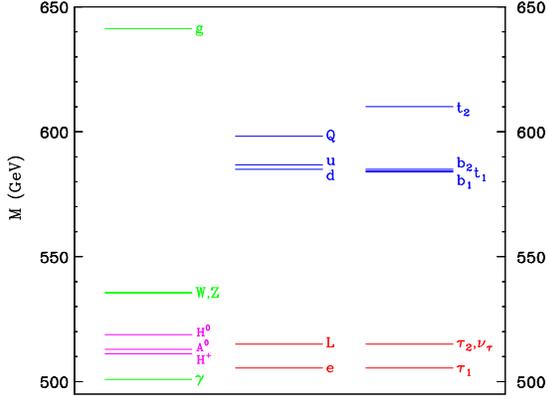}
%\framebox[55mm]{\rule[-21mm]{0mm}{43mm}}
\caption{One-loop corrected mass spectrum of
the first KK level in UED for $R^{-1}=500$ GeV
and SM Higgs mass $m_h=120$ GeV. From Ref.~\cite{Cheng:2002iz}.}
\label{fig:uedspectrum}
\end{figure}

\begin{figure}[tb]
\includegraphics[width=17pc]{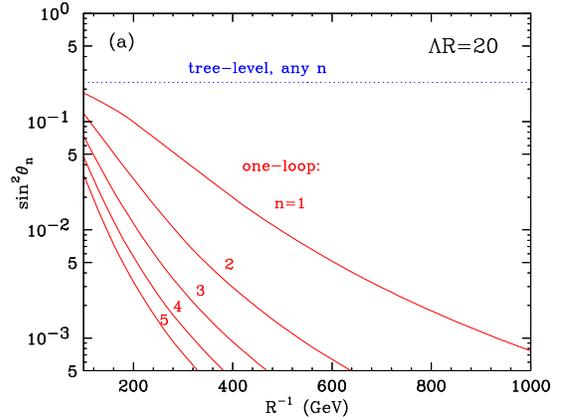}
%\framebox[55mm]{\rule[-21mm]{0mm}{43mm}}
\caption{The Weinberg angle $\theta_n$ for the first few KK levels
as a function of $R^{-1}$. From Ref.~\cite{Cheng:2002iz}.}
\label{fig:s2t}
\end{figure}

Including the one-loop radiative corrections, 
the mass spectrum of the level 1 KK-partners
(see Fig.~\ref{fig:uedspectrum}) 
exhibits striking similarities to supersymmetry
\cite{Cheng:2002ab}.
The LKP is a neutral WIMP which is a linear combination of the 
first KK mode $B_1$ of the hypercharge
gauge boson and the first KK mode $W^0_1$ of the neutral 
$SU(2)_W$ gauge boson. The mass matrix for the neutral gauge bosons
at level $n$ has the form
$$
\left( 
\begin{array}{cc}
\frac{n^2}{R^2}
+ \frac{1}{4}g_1^2 v^2 
+ \hat{\delta} m_{B_n}^2 
& \frac{1}{4}g_1 g_2 v^2 \\
\frac{1}{4}g_1 g_2 v^2 & 
\frac{n^2}{R^2}
+\frac{1}{4}g_2^2 v^2
+ \hat{\delta} m_{W_n}^2 
\end{array}
\right)\ .
$$
Upon diagonalization, we find the Weinberg angle $\theta_n$ 
at KK level $n$
%\begin{equation}
%\gamma_n = \cos\theta_n B^0_n + \sin\theta_n W^0_n \approx B^0_n.
%\end{equation}
shown in Fig.~\ref{fig:s2t}. We see that while at tree level $\theta_n$
is identical to the SM value, at one loop it is rather small.
As a result, the LKP is predominantly $B_1$, in analogy to the case of 
$\tilde b^0$ in SUSY.

\subsection{Relic density of KK dark matter}

\begin{figure}[htb]
\includegraphics[width=18pc]{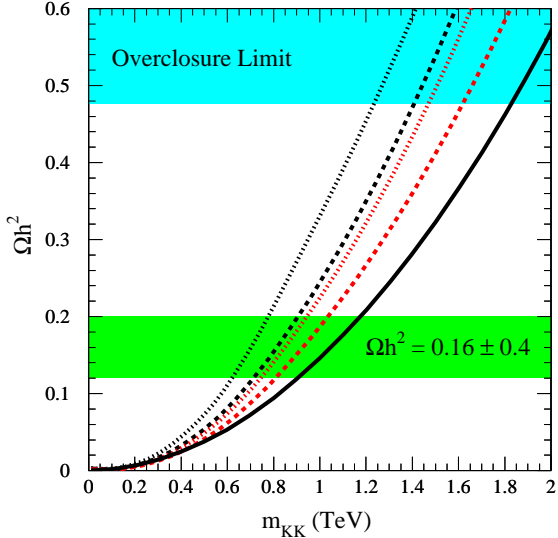}
%\framebox[55mm]{\rule[-21mm]{0mm}{43mm}}
%\vspace*{-1.0cm}
\caption{Prediction for $\Omega_{CDM}h^2$ in UED as a function of
the LKP mass $M_{KK}\sim R^{-1}$. The solid line is the case of no 
coannihilations while the dashed (dotted) lines are for
the case of coannihilations with one (three) generations 
of KK leptons and different degrees of mass degeneracy.
From Ref.~\cite{Servant:2002aq}.}
\label{fig:uedrd}
\end{figure}

The relic density of Kaluza-Klein dark matter
can be readily computed~\cite{Servant:2002aq}.
Unlike the case of supersymmetry, where the LSP is a Majorana fermion, 
here the LKP is a vector particle and the helicity suppression
is absent. The LKP relic density
\begin{equation}
\Omega h^2 = \frac{1.04\ 10^{9}\ {\rm GeV}^{-1}}{M_P\sqrt{g_\ast}}
\frac{x_F}{a+3b/x_F}
\end{equation}
is mostly determined by the $a$-term in the velocity expansion of 
the annihilation cross-section, where~\cite{Servant:2002aq}
\begin{equation}
a = \frac{\alpha_1^2}{M_{KK}^2}\frac{380\pi}{81}; 
\qquad 
b = -\frac{\alpha_1^2}{M_{KK}^2}\frac{95\pi}{162}.
\end{equation}
Here $M_{KK}\sim R^{-1}$ is the LKP mass, 
$x_F^{-1}=T_F/M_{KK}$ is the dimensionless freeze-out temperature and
$\alpha_1$ is the hypercharge gauge coupling constant.
Unlike supersymmetry, coannihilation processes in UED
{\em lower} the preferred LKP mass range (see Fig.~\ref{fig:uedrd}).

\subsection{Detection of KK dark matter}

\begin{figure}[htb]
\includegraphics[width=18pc]{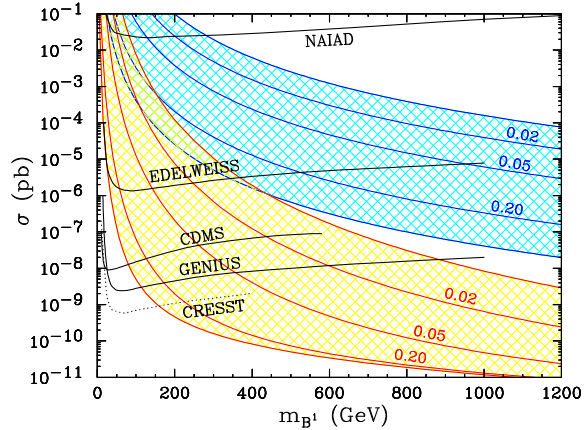}
%\framebox[55mm]{\rule[-21mm]{0mm}{43mm}}
\caption{Range of predicted spin-dependent proton cross sections (dark-shaded),
along with the projected sensitivity of a 100 kg NAIAD
array~\cite{Spooner:kt}; and predicted spin-independent proton cross
sections (light-shaded), along with the current EDELWEISS
sensitivity~\cite{Benoit:2002hf}, and projected sensitivities of
CDMS~\cite{Schnee:gf}, GENIUS~\cite{Klapdor-Kleingrothaus:2000eq}, and
CRESST~\cite{Bravin:1999fc}. From Ref.~\cite{Cheng:2002ej}.
}
\label{fig:ueddirect}
\end{figure}

The prospects for direct detection of KK dark matter are 
summarized in Fig.~\ref{fig:ueddirect}~\cite{Cheng:2002ej}. 
As in the case of SUSY,
only spin-independent probes appear promising, since
precision data has ruled out $R^{-1}<250$ GeV~\cite{Appelquist:2002wb}.
The predictions are for $m_h = 120$ GeV and $0.01 \le r =
(M_{Q_1} - M_{B_1}) / M_{B_1} \le 0.5$, 
with contours for specific intermediate $r$ labeled.
A nice feature of UED is the constructive interference of 
the relevant diagrams, guaranteeing a 
lower bound on the detection rate, which is 
evident in Fig.~\ref{fig:ueddirect}.

\section{LITTLE HIGGS THEORIES}
\label{sec:LH}

\begin{figure}[htb]
\begin{center}
{
%\unitlength=1.0 pt
%\SetScale{1.0}
%\SetWidth{0.7}      % line    size control
\unitlength=0.7 pt
\SetScale{0.7}
\SetWidth{0.7}      % line    size control
\footnotesize    %  letter  size control
{} %\qquad
\allowbreak
%  diagram # 1
\begin{picture}(300,100)(0,0)
\SetColor{Red}
\DashLine(10,20)(90,20){3}
\DashCArc(50,50)(30,0,360){3}
\DashLine(110,20)(190,20){3}
\PhotonArc(150,50)(30,0,360){3}{15}
\DashLine(210,50)(230,50){3}
\DashLine(290,50)(310,50){3}
\CArc(260,50)(30,0,360)
\Text( 50,50)[c]{Higgs}
\Text(150,50)[c]{$W,Z,\gamma$}
\Text(260,50)[c]{Top}
\Text( 12,27)[c]{$h$}
\Text( 88,27)[c]{$h$}
\Text(112,27)[c]{$h$}
\Text(188,27)[c]{$h$}
\Text(212,57)[c]{$h$}
\Text(308,57)[c]{$h$}
\Text( 50,10)[c]{$\lambda$}
\Text(150,10)[c]{$g^2$}
\Text(225,40)[c]{$\lambda_t$}
\Text(298,40)[c]{$\lambda_t$}
\end{picture}
{} %\qquad
\allowbreak
%  diagram # 1
\begin{picture}(300,100)(0,0)
\SetColor{Red}
\DashLine(10,20)(90,20){3}
\DashLine(110,20)(190,20){3}
\DashLine(210,20)(290,20){3}
\SetColor{Blue}
\DashCArc(50,50)(30,0,360){3}
\PhotonArc(150,50)(30,0,360){3}{15}
\CArc(250,50)(30,0,360)
\Vertex(250,80){3}
\Text( 50,50)[c]{$H'$}
\Text(150,50)[c]{$W',Z',\gamma'$}
\Text(232,50)[c]{$\chi_L$}
\Text(270,50)[c]{$\chi_R$}
\Text( 12,27)[c]{$h$}
\Text( 88,27)[c]{$h$}
\Text(112,27)[c]{$h$}
\Text(188,27)[c]{$h$}
\Text(212,27)[c]{$h$}
\Text(288,27)[c]{$h$}
\Text( 50,10)[c]{$-\lambda$}
\Text(150,10)[c]{$-g^2$}
\Text(250,90)[c]{$\lambda_t f$}
\Text(250,10)[c]{$-\lambda_t/(2f)$}
\end{picture} 
}
% END OF DIAGRAMS
\end{center}
\caption{Top: SM diagrams leading to a quadratic divergence in the
one-loop Higgs mass correction. Bottom: New physics contributions
from Little Higgs theories which cancel the quadratic divergence.}
\label{fig:SMdivergence}
\end{figure}
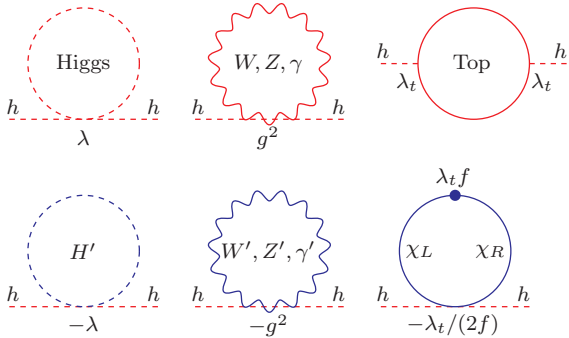

Little Higgs theories alleviate the Higgs hierarchy problem 
by eliminating the quadratic divergence in the 
Higgs mass correction at one-loop (Fig.~\ref{fig:SMdivergence}).
New particles are introduced around the TeV scale 
and their couplings are such that the quadratic divergence 
cancels for each vertical pair of diagrams in Fig.~\ref{fig:SMdivergence}.
The special values of the couplings are guaranteed by a
symmetry, as the Higgs boson is a pseudo-Goldstone boson.

However, generic Little Higgs models give tree-level 
contributions to precision data and are severely 
constrained~\cite{Csaki:2002qg}.
One possible solution is to postulate a 
conserved $T$-parity \cite{Cheng:2003ju},
which may have a geometrical origin in theory space.
The SM particles are chosen to be even under $T$-parity 
while all new particles are odd.
This ensures that the lightest $T$-odd particle (LTP) is stable
and again may serve as a DM candidate.

\subsection{Dark matter in Little Higgs models}

\begin{figure}[htb]
\includegraphics[width=17pc]{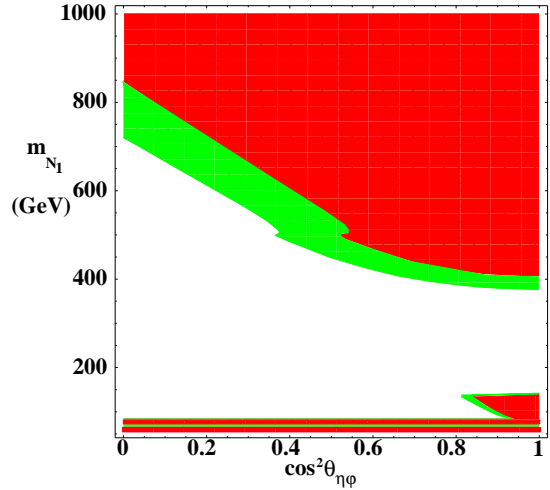}
%\framebox[55mm]{\rule[-21mm]{0mm}{43mm}}
\caption{Preferred relic density regions (green, light shaded) 
as a function of $\cos\theta_{\eta\varphi}$ and $m_{N_1}$.
From Ref.~\cite{Birkedal-Hansen:2003mp}.}
\label{fig:ScalarDM}
\end{figure}

The relic density of little Higgs dark matter was studied in
\cite{Birkedal-Hansen:2003mp} for a specific model with a 
scalar LTP $N_1$, which is
a mixture of an SU(2)-triplet $\varphi^0_{\bf 3}$ and
an SU(2)-singlet $\eta^0_{\bf 1}$ (see Fig.~\ref{fig:ScalarDM})
\begin{equation}
N_1 = \cos\theta_{\eta\varphi}\, \eta^0_{\bf 1}
    + \sin\theta_{\eta\varphi}\, \varphi^0_{\bf 3}\ .
\end{equation}
The absence of a helicity suppression requires
relatively large masses $m_{N_1}$ for the LTP WIMP case.
However, for $150\ {\rm GeV}<m_{N_1}<350\ {\rm GeV}$, 
annihilation into $t\bar{t}$ and  $hh$ is very efficient
(even if $N_1$ is a pure singlet) and there is no 
preferred region for any value of the mixing angle 
$\theta_{\eta\varphi}$.

\section{SUSY-UED DISCRIMINATION}
\label{sec:colliders}

We saw that the superpartners and level 1 KK partners in UED 
may have a very similar spectrum. In addition, both models 
have a DM candidate
which could give a signal in upcoming dark matter detection 
experiments. The natural question then is, how can one distinguish
the two scenarios? The two basic methods are: producing and studying
the dark matter at a high energy collider; or looking for 
certain indirect DM signals~\cite{JF}.

Recall that the mass spectrum at each KK level in UED is
rather degenerate (Fig.~\ref{fig:uedspectrum}).
The typical UED collider signatures therefore include
soft leptons, soft jets and a modest amount of $\met$.

\subsection{Hadron colliders}

\begin{figure}[htb]
\includegraphics[width=17pc]{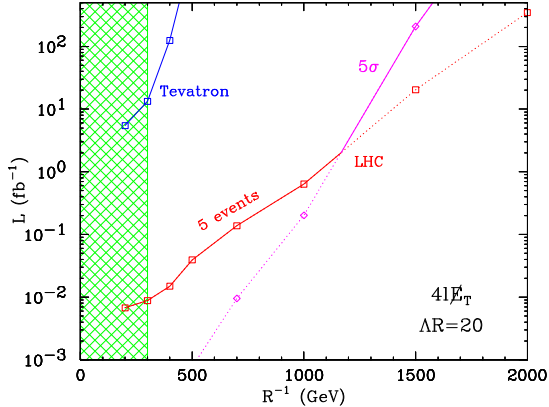}
%\framebox[55mm]{\rule[-21mm]{0mm}{43mm}}
\caption{UED discovery reach at
the Tevatron (blue) and the LHC (red) in the $4\ell\met$ channel. 
We require a $5\sigma$ excess or the observation of 5 signal events,
and show the required total integrated luminosity per experiment 
(in ${\rm fb}^{-1}$) as a function of $R^{-1}$. From Ref.~\cite{Cheng:2002ab}.}
\label{fig:uedLHC}
\end{figure}

At hadron colliders one would like to use the strong production
of KK-quarks and KK-gluons, followed by a decay chain through
$W^\pm_1$ and $W^0_1$, whose decays always yield leptons and/or 
neutrinos (see Fig.~\ref{fig:uedspectrum}).
Inclusive pair production of $SU(2)_W$ doublet quarks $Q_1$
results in a clean $4\ell\met$ inclusive signature 
with a reasonable rate.
This channel was studied in~\cite{Cheng:2002ab} for both the 
Tevatron and the LHC and the resulting reach is shown in 
Fig.~\ref{fig:uedLHC}.
We see that for nominal integrated luminosities 
the LHC will see a UED signal throughout all of the
cosmologically preferred range (see Fig.~\ref{fig:uedrd}).
However, UED and supersymmetry can be easily confused 
at hadron colliders, where spin determinations 
are rather challenging because of the unknown 
parton-level center-of-mass energy in the event. 

\subsection{Linear colliders}

\begin{figure}[htb]
\includegraphics[width=17pc]{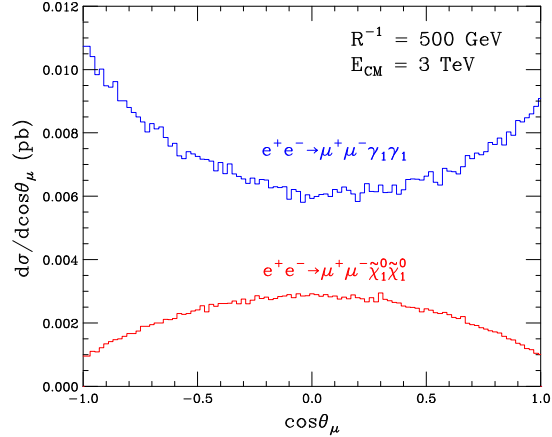}
%\framebox[55mm]{\rule[-21mm]{0mm}{43mm}}
\caption{Differential cross-section $d\sigma/d\cos\theta_\mu$
for $\mu_1$ production (\ref{KKmuon}) in UED (blue, top) and 
$\tilde\mu$ production (\ref{smuon}) in supersymmetry (red, bottom), 
as a function of the muon scattering angle $\theta_\mu$.
We have chosen $R^{-1}=500$ GeV in UED
and then adjusted the SUSY breaking parameters until
we get a perfect spectrum match~\cite{BDKMR}.}
\label{fig:susy_ued}
\end{figure}

In contrast, at linear colliders the information 
about the spin of the produced particle is
encoded in the angular distributions of its decay products.
Consider, for example the two similar scenarios of 
KK-muon production in UED
\begin{equation}
e^+e^-\to \mu^+_1\mu^-_1\to \mu^+\mu^- \gamma_1 \gamma_1
\label{KKmuon}
\end{equation}
and smuon production in supersymmetry
\begin{equation}
e^+e^-\to \tilde\mu^+\tilde\mu^-
\to \mu^+\mu^-\tilde\chi^0_1\tilde\chi^0_1\ .
\label{smuon}
\end{equation}
The angular distribution of the KK-muon $\mu_1$
(with respect to the beam axis) is given by
\begin{equation}
\frac{d\sigma}{d\cos\theta} \sim 1+\cos\theta^2\ ,
\end{equation}
while for the smuon $\tilde\mu$ we have
\begin{equation}
\frac{d\sigma}{d\cos\theta} \sim 1-\cos\theta^2\ .
\end{equation}
In practice one measures the {\em muon} scattering angle $\theta_\mu$,
which is significantly different for the two cases at hand 
(Fig.~\ref{fig:susy_ued}).
In addition, the spin difference is reflected in
the total cross-section, which can be measured very well,
as well as in the energy dependence of the total cross-section 
at threshold.

\section{CONCLUSIONS}
\label{sec:conclusions}

Table~\ref{table:summary} summarizes the main features of
the models discussed in Sections~\ref{sec:susy}-\ref{sec:LH}.
\begin{table*}[htb]
\caption{Summary of the models discussed in the text.}
\label{table:summary}
\newcommand{\m}{\hphantom{$-$}}
\newcommand{\cc}[1]{\multicolumn{1}{c}{#1}}
\renewcommand{\tabcolsep}{2pc} % enlarge column spacing
\renewcommand{\arraystretch}{1.2} % enlarge line spacing
\begin{tabular}{lccc}
\hline\hline
%Higgs & \multicolumn{3}{c||}{Higgs production mechanisms} \\ 
%\cline{2-4}
%decays
Model
& SUSY
& UED
& Little Higgs \\ 
\hline\hline
DM particle
& LSP
& LKP
& LTP \\ 
\hline
Spin
& 1/2
& 1
& 0 \\ 
\hline
Symmetry
& $R$-parity
& KK-parity
& $T$-parity \\ 
\hline
DM mass range
& 50-200 GeV 
& 600-800 GeV
& 400-800 GeV \\ 
\hline
\hline
\end{tabular}
\end{table*}
By now there are several well-motivated alternatives for BSM WIMPs
and discriminating among them will be a challenge for
the upcoming direct detection and hadron collider experiments.
In this sense, astroparticle physics experiments in space may 
play an important role as discriminators among various DM alternatives
and thus provide important clues to the identity of the dark matter.

\end{document}